\documentstyle[12pt]{article}
\newcommand{\doublespacing}{\let\CS=\@currsize\renewcommand{\baselinesstrech}
{2.0}\tiny\CS}

\begin{document}

\title{Time dependent nonclassical properties of even and odd nonlinear
coherent states}
\author{B.Roy.\thanks{E-mail : barnana@isical.ac.in}\\ and\\
P. Roy\thanks{E-mail : pinaki@isical.ac.in}\\ Physics \& Applied
Mathematics Unit\\ Indian Statistical Institute \\ Calcutta  700035\\ India} 

\date{}

\maketitle

\vspace*{1.5cm}

\centerline{\bf Abstract}

\vspace{0.3cm}

\thispagestyle{empty}

\setlength{\baselineskip}{18.5pt}

We construct even and odd nonlinear coherent states of a parametric oscillator
and examine their nonclassical properties.It has been shown that these superpositions
exhibit squeezing and photon antibunching which change with time. 

\newpage
\section{Introduction}

Coherent states of various Lie algebras as well as various superpositions of
coherent states have attracted considerable attention over the years.It has been
shown that particular superpositions of harmonic oscillator coherent states 
known as the even and odd coherent states [1],exhibit nonclassical properties like
squeezing and antibunching [2-4].Recently another type of coherent states,called
the f - coherent states [5] or the nonlinear coherent states [6] has been 
introduced.It has been shown [6] that nonlinear coherent states with a particular
class of nonlinearities is useful in the description of a trapped ion and that
such states have strong nonclassical properties.Subsequently various superpositions
of nonlinear coherent states have been studied [7,8] and shown to possess nonclassical
properties like squeezing,antibunching etc.Recently a scheme of generating even and
odd nonlinear coherent states has also been suggested [9].

On the other hand time dependent quantum systems are of interest in different
areas of physics including quantum optics.In particular time dependent harmonic
oscillator has been widely studied and exact solutions of a class of parametric
oscillator with time dependent freequency is known [10].Using these solutions one
can construct coherent states [11] as well as various superpositions of coherent
states e.g, even and odd coherent states [12].It has been shown that these cat
states possess interesting nonclassical properties which depend on time [12].In the
present work our objective is to examine nonclassical properties of even and odd
nonlinear coherent states of a parametric oscillator.More specifically we shall
study squeezing and antibunching properties of these states and examine how these
properties change with time.

\section{Even and odd nonlinear coherent states of a parametric oscillator}

Here we shall consider a time dependent oscillator with the Hamiltonian (we take
$h = 2m = \omega(0) = 1$)

\begin{equation}
H = - \frac{1}{2}\partial^2 / \partial x^2 +
\frac{1}{2}\omega^2(t)x^2 
\end{equation}

The time dependent integral of motion are then given by [10]

\begin{equation}
A = \frac{i}{\sqrt2}[\epsilon(t)p - \dot{\epsilon}(t)x]
\end{equation}

where $\epsilon(t)$ satisfies the following conditions :

\begin{equation}
\ddot{\epsilon}(t) + \omega^2(t)\epsilon(t) = 0 ,~~~\epsilon(0) = 1,~~~\dot\epsilon(0) = i
\end{equation}

It can be readily verified that the operators $A$ and $A^{\dagger}$ satisfy the
commutation relation

\begin{equation}
            [A , A^{\dagger}] = 1
\end{equation} 

It can be shown [10] that the normalised eigenstates of the Schroedinger equation
corresponding to (1) is given by

\begin{equation}
\psi_n(x,t) = (\frac{\epsilon^*(t)}{2\epsilon(t)})^{\frac{n}{2}}\frac{1}{\sqrt(n!)}\psi_0(x,t)H_n(\frac{x}{|\epsilon(t)|})  
\end{equation}

where $\psi_{0}(x,t)$ is given by

\begin{equation}
\psi_0(x,t) = \pi^{-\frac{1}{4}}\epsilon(t)^{-\frac{1}{2}}exp(\frac{i\dot\epsilon(t)x^2}{2\epsilon(t)})
\end{equation}

The coherent state corresponding to the parametric oscillator (1) can now be 
constructed as in the time independent case and it is given by

\begin{equation}
\psi_{\beta}(x,t) = \sum_{n=0}^{\infty}\frac{\beta^n}{n!} \psi_n(x,t) =
\psi_0(x,t)exp(-\frac{1}{2}|\beta|^2 -
\frac{\beta^2\epsilon^*(t)}{2\epsilon(t)} + \frac{\sqrt{2}\beta x}{\epsilon(t)})
\end{equation}

where $\beta$ is an arbitrary complex number.It can now be easily verified that
the coherent states (7) are also eigenstates of the operator $A$:

\begin{equation}
                 A \psi_{\beta}(x,t) = \beta \psi_\beta(x,t)
\end{equation}

We now proceed to the construction of time dependent even and odd nonlinear
coherent states.For the sake of convenience let us denote the time dependent
coherent state by the ket $|\beta,t>$ so that $<x|\beta,t>$ i.e,the wave function representation of
the coherent state is given by equation (7).First we note that the time 
dependent nonlinear coherent states are eigenstates of a generalised time
dependent integral of motion B defined by

\begin{equation}
                B = f(A^{\dagger}A)A
\end{equation}

where $f(A^{\dagger}A)$ is a real function and is called the nonlinearity function.Thus the generalised
integrals of motion $B$ and $B^{\dagger}$ satisfy the following algebra:

\begin{equation}
[B , B^{\dagger}] =  f^2(A^{\dagger}A) (A^{\dagger}A+1) - f^2(A^{\dagger}A-1) A^{\dagger}A
\end{equation}

Thus in contrast to $A$ and $A^{\dagger}$ the operators $B$ and $B^{\dagger}$
satisfy a deformed Heisenberg algebra whose nature of deformation depends on
on the choice of the nonlinearity function $f$.Clearly if $f=1$ the relation
(10) becomes the same as (4).Now defining the time dependent nonlinear coherent
state $|\alpha,t>_{NL}$ as right eigenstate of the oprator B we obtain using (9)

\begin{equation}
|\alpha,t>_{NL}~~=~~C \sum_{n=0}^{\infty}d_n{\alpha}^n |n,t>
\end{equation}

where $\alpha$ is a complex number and $|n,t>$ denotes the ket corresponding to the wave function (5).The normalisation
constant C and the coefficients $d_n$ are given by

\begin{equation}
\begin{array}{lcl}
C^2& =& \left[ \sum_{n=0}^{\infty} d_n^2 (\bar {\alpha} \alpha)^n \right]^{-1}\\
d_0 &=& 1 \\
d_n &=& \left[ \sqrt{n!}f(n)!\right]^ {-1}
\end{array}
\end{equation}
where $f(n)! = f(1)f(2)....f(n)$.

Even nonlinear coherent states of a parametric oscillator are then defined by

\begin{equation}
|\alpha,t>_{ENL}~~=~~N_{+} ( |\alpha,t>_{NL}~~+~~|-\alpha,t>_{NL} )= C_{+}
\sum_{n=0}^{\infty} d_{2n} |2n,t> 
\end{equation} 

where $C_{+}$ is a normalisation constant and is given by
\begin{equation}
C_{+} = [\sum_{n = 0}^{\infty} d_{2n}^2 |\alpha|^{4n}]^{-1/2}
\end{equation}

Proceeding similarly the odd nonlinear coherent states are found to be

\begin{equation}
|\alpha,t>_{ONL}~~=~~N_{-} ( |\alpha,t>_{NL}~~-~~|-\alpha,t>_{NL} ) = C_{-}
\sum_{n=0}^{\infty} d_{2n+1}{\alpha^{2n+1}} |2n+1,t> 
\end{equation}

where $C_{-}$ is given by

\begin{equation}
C_{-} = [\sum_{n = 0}^{\infty} d_{2n+1}^2 |\alpha|^{4n+2}]^{-1/2}
\end{equation}

\section{Squeezing of the even nonlinear coherent states}
Before discussing squeezing we note that the standard time independent harmonic oscillator creation and annihilation operators
$a$ and $a^{\dagger}$ are connected to the time dependent one $A$ and $A^{\dagger}$
by the following relation :
\begin{equation}
\left( \begin{array}{l}   a \\ 
              a^{\dagger}\\ \end{array} \right) ~=~  \left( \begin{array}{lcl}
                                            u^{\ast} & -v \\ 
 -v^{\ast} & u \\ \end{array} \right) ~ \left( \begin{array}{l}  A \\ 
                   A^{\dagger}\\ \end{array} \right) 
\end{equation}

where $u$ and $v$ are defined in terms of $\epsilon$ and $\dot\epsilon$ :

\begin{equation}
u = \frac{1}{2}(\epsilon - i\dot\epsilon),~~~v = - \frac{1}{2}(\epsilon + i\dot\epsilon)
\end{equation}

Thus the relations (17) enables us to determine the action of the operators $a$
and $a^{\dagger}$ on the states given by (13) and (15).

Now to examine squeezing behaviour of the even and odd nonlinear coherent states
we introduce the following quadratures :

\begin{equation}
           X_1 = \frac{a + a^{\dagger}}{\sqrt{2}},~~~ X_2 = \frac{a - a^{\dagger}}{\sqrt{2}i}
\end{equation}

Then it follows that the operators $X_1$ and $X_2$ satisfy the following uncertainty
relation:

\begin{equation}
<\Delta X_1^2>~~<\Delta X_2^2>~~~\geq~~~\frac{1}{4}~<[X_1,X_2]>^2
\end{equation}

Where $<\Delta X_i^2>$ is defined as

\begin{equation}
<\Delta X_i^2>~=~ <X_i^2> - <X_i>^2
\end{equation}
 
Thus it is clear that a state is squeezed if either of the following inequalties hold:

\begin{equation}
< \Delta X_1^2>~~~<~~~\frac{1}{2}~|<[X_1,X_2]>|~~~or~~~<\Delta X_2^2>~~~<~~~\frac{1}{2}~|<[X_1,X_2]>|
\end{equation}

Now to examine whether or not the squeezing conditions (22) hold we need to
evaluate several expectation values like $<A^{\dagger}A>$,$<A^2>$,$<A^{\dagger^2}>$
etc.This in turn requires a specific choice of the parameter $\epsilon(t)$.For
a given freequency $\omega(t)$ this can be obtained from the solution of equation (3).Here we take the oscillator
freequency $\omega(t)$ and the corresponding parameter $\epsilon(t)$ as [12,13]

\begin{equation}
\begin{array}{lcl}
\omega(t)& =&\displaystyle{ \frac{1+\kappa cos(2t)}{1+\kappa}}\\ \\
\epsilon(t)& =&\displaystyle{ cosh(\frac{1}{4}\kappa t)e^{it} + isinh(\frac{1}{4}\kappa
t)e^{-it}~~~,~~~\kappa << 1}\\
\end{array}
\end{equation}

It may be pointed out that with this choice of $\epsilon(t)$,the functions
$u$ and $v$ defined in (18) become determined too.Finally we need to specify the nonlinearity function $f$.Obviously for each
choice of the nonlinearity function we get a different nonlinear coherent state.
In the present case we take as nonlinearity fuction the one considered in ref [6]
to describe the motion of a trapped ion :

\begin{equation}
f(n) = L_n^1(\eta^2)[(n+1)L_n^0(\eta^2)]^{-1} 
\end{equation}

where $L_n^m(x)$ are generalised Laguerre polynomials and $\eta$ is known
as the Lamb-Dicke parameter.Clearly $f(n)=1$ when $\eta=0$ and in this case the even and odd
nonlinear coherent states become standard even and odd coherent states.However as
$\eta$ increases the nonlinearity develops and it is reflected by the structure
of the phase probability distribution [14] of the even and odd
nonlinear coherent states.It has been shown that for a reasonably large value
of the Lamb-Dicke parameter $\eta$/$\alpha$ the phase probability
distributions have multiple peaks due self splitting(which is a consequence of
nonlinearity).For very large values of the Lamb-Dicke parameter $\eta$/$\alpha$ the
number of peaks also increases and the structure becomes fairly complicated.
So in what follows we shall confine ourselves to reasonable values of $\eta$
and $\alpha$ such that all essential features associated with nonlinearity are present
without being too complicated. 

It may be noted that with this choice of the nonlinearity function the algebra
(10) becomes a nonpolynomial deformation of the Heisenberg algebra.We now proceed
to examine the inequalities in (22).After some calculations using (17) and (19) as well as (23) and
(24) it can be shown that the uncertainty relation (20) and
the squeezing conditions (22) can be expressed respectively as

\begin{equation}
F(\alpha,\eta,\kappa,t)~G(\alpha,\eta,\kappa,t)~~~\geq~~~\frac{1}{4}
\end{equation}
and
\begin{equation}
F(\alpha,\eta,\kappa,t)~~<~~\frac{1}{2}~~or~~G(\alpha,\eta,\kappa,t)~~<~~\frac{1}{2}
\end{equation}

where $F(\alpha,\eta,\kappa,t)$ and $G(\alpha,\eta,\kappa,t)$ are given by

\begin{equation}
\begin{array}{lcl}
F(\alpha,\eta,\kappa,t)& = & \frac{2}{(\kappa cos(2t) + 4)}(\epsilon^2 <A^{\dagger^2}> +
\epsilon^{*^2} <A^2> + 2\epsilon\epsilon^* <A^{\dagger}A> + \epsilon\epsilon^*)\\
G(\alpha,\eta,\kappa,t)& = & \frac{2}{(\kappa cos(2t) + 4)}(\dot\epsilon^2 <A^{\dagger^2}
+ \dot\epsilon^{*^2} <A^2> - 2\dot\epsilon^*\dot\epsilon <A^{\dagger}A> - \dot\epsilon^*\dot\epsilon)\\
\end{array}
\end{equation}
and in obtaining the above expressions we have neglected terms of higher order
in $\kappa$.

We now proceed to describe the figures.In figs 1 and 2 we describe the plots of $F(\alpha,\eta,\kappa,t)$
and $G(\alpha,\eta,\kappa,t)$ against time $t$ for the even nonlinear coherent state.
From figure 1 we find that the even nonlinear coherent state exhibits squeezing
in both quadratures at different times.From figure 1 we also find that 
at some instant of time $F(\alpha,\eta,\kappa,t) = G(\alpha,\eta,\kappa,t) =.5$
so that the uncertainty relation (25) is saturated.This implies that even
nonlinear coherent states can also be instantaneous intelligent states. 
In fig 2 we plot the same quantities as in fig 1
but for different parameter values.From fig 2 we find that squeezing behaviour
remains the same as before.However comparing fig 1 and fig 2 it is seen 
that squeezing increases as $\alpha$ increases.
 
Now a word about squeezing behaviour of the odd nonlinear coherent states.We have
evaluated $F(\alpha,\eta,\kappa,t)$ and $G(\alpha,\eta,\kappa,t)$ over a wide range of values
of $\alpha$ and $\eta$ but it was found that they never satisfy the inequalities
in (24).Thus we conclude that the odd nonlinear coherent state does not exhibit
quadrature squeezing.

\section{Antibunching of the odd nonlinear coherent states}

In order to study antibunching property we have to evaluate the second order
correlation function $g^2(0)$ and the condition for antibunching is given by

\begin{equation}
g^2(0) = \frac{<a^{\dagger^2}a^2>}{<a^{\dagger}a>^2}~~<~~1
\end{equation}

Now using equation (17) we obtain

\begin{equation}
\begin{array}{lcl}
<a^{\dagger^2}a^2>& = & |u|^2|v|^2 + 4|u|^2|v|^2 <A^{\dagger}A> +
4|u|^2|v|^2 <(A^{\dagger}A)^2> + |u|^4 <A^{\dagger^2}A^2>\\ 
&&+ |v|^4 <A^2A^{\dagger^2}> - \left[(u^2u^{*}v <A^{\dagger^2}> +
u^{*}vv^{*^2} <A^2> + 2uv^{*}u^{*^2} <A^{\dagger}A^3> \right.\\
&&\left. + 2u^{*}vv{*^2} <A^2A^{\dagger}A> 
- u^2v^2<A^{\dagger^4}>) + c.c \right]\\
<a^{\dagger}a>& = & (|u|^2 + |v|^2)<A^{\dagger}A> - |v|^2 - u^*v^*<A^2> - uv <A^{\dagger^2}>\\
\end{array}
\end{equation}

We have evaluated $g^2(0)$ for the odd nonlinear coherent state and the result
is plotted in figure 3 against time $t$.From figure 3 we find that while $g^2(0)$
has an increasing trend it is less than $1$ for a considerable perod of time.
This implies that the odd nonlinear coherent states exhibit antibunching.The plot
of $g^2(0)$ shows that it has an increasing trend and eventually for large values
of time it may become more than $1$.

Finally we note that for the even nonlinear coherent state we have evaluated
$g^2(0)$ for a large number of values of the parameters but it was found to very
much larger than unity.We thus conclude that the even nonlinear coherent state
does not exhibit antibunching at any time.

{\bf Conclusion}

In this article we have studied squeezing and antibunching properties of the
even and the odd nonlinear coherent states of a parametric oscillator.It has 
been shown that while the squeezing of the even nonlinear coherent states
increases with time,antibunching of the odd nonlinear coherent states decreases
with time.

\newpage

\noindent
{\large{\bf References}}

\vspace{0.5cm}

\begin{enumerate}
\item [[1]] V.V.Dodonov,I.A.Malkin and V.I.Man'ko, Physica 72 (1972) 597
\item[[2]] M.Hillery, Phys.Rev A36 (1987) 3796
\item[[3]] Y.Xia and G.Guo, Phys.Lett A136 (1989) 281
\item[[4]] C.C.Gerry, J.Mod.Opt 40 (1993) 1053
\item[[5]]V.I.Man'ko, G.Marmo, E.C.G.Sudarshan and F.Zaccaria, Physica Scripta 
55 (1997) 528
\item[[6]] R.L.de Matos and W.Vogel, Phys.Rev A54 (1996) 4560
\item[[7]] S.Mancini, Phys.Lett A233 (1997) 291
\item[[8]] B.Roy, Phys.Lett A249 (1998) 25
\item[[9]] S.Sivakumar, preprint,quant-ph/9902054
\item[[10]] I.A.Malkin and V.I.Man'ko, Phys.Lett A32 (1970) 243
\item[[11]] I.A.Malkin,V.I.Man'ko and D.A.Trifonov, Phys.Rev D2 (1970) 1371
\item[[12]] N.A.Ansari and V.I.Man'ko, Phys.Lett A223 (1996) 31
\item[[13]] V.V.Dodonov,V.I.Man'ko and D.E.Nikonov, Phys.Rev A51 (1995) 3328
\item[[14]] B.Roy and P.Roy, Phys.Lett A257 (1999) 264
\end{enumerate}
\end{document}